\documentclass[a4paper,11pt]{article}
\usepackage{jheppub}
\usepackage{bookmark}
\usepackage{array}

\usepackage{amsmath}
\usepackage{mathtools}
\usepackage{amssymb}
\usepackage{wrapfig}
\usepackage{xspace}
\usepackage{float}
\usepackage{pgfplots}
\usepackage{pgfplotstable}
\usepackage{ifthen}
\usepackage[utf8]{inputenc}

\usepackage[compat=1.1.0]{tikz-feynman}

\newcommand{\rhoVar}[3][]{
    \ensuremath{
        \mathbf{\rho}
        \ifthenelse{\equal{#1}{}}{\!}{^#1 \!}
        \left( #2;#3 \right )
    }
}

\renewcommand{\Im}{\mathrm{Im}}

\newcommand{\UV}{\mathrm{UV}}
\newcommand{\IR}{\mathrm{IR}}

\newcommand{\maxN}{\texttt{maxN}\xspace}
\newcommand{\maxJ}{\texttt{maxJ}\xspace}

\newcommand{\aFree}{a_{\textrm{free}}}

\definecolor{plot1}{HTML}{2a70ad}
\definecolor{plot2}{HTML}{3c9974}
\definecolor{plot3}{HTML}{c9651d}
\definecolor{plot4}{HTML}{9f3f74}

\pgfplotsset{compat=1.3}

\arxivnumber{2307.02305}

\title{Bootstrapping the $a$-anomaly in $4d$ QFTs: Episode II}

\author{J. K. Marucha}
\affiliation{Fields and Strings Laboratory, Institute of Physics
École Polytechnique Fédérale de Lausanne (EPFL)
Route de la Sorge, CH-1015 Lausanne, Switzerland}
\emailAdd{jan.marucha@epfl.ch}

\abstract{
As recently shown \cite{a_anomaly_paper},  the a-anomaly of the UV fixed point of 4d quantum field theories, can be constrained by studying scattering amplitudes.
The basic idea is to couple the QFT to a  dilaton and impose unitarity of the scattering amplitudes of physical particles and probe dilatons.
In this work we find new lower bounds on the $a$-anomaly in several  gapped QFTs containing different numbers of stable scalar particles.}

\begin{document}
\maketitle
\section{Introduction and summary}\label{introduction}

The $a$-anomaly coefficient of 4d CFTs is of great interest to physicists. Not only it can be thought as, analogous to the 2d central charge $c$, as a way of measuring the number of degrees of freedom, but also, exactly like $c$ for 2d, it is strictly monotonic along RG flow, with $a_{UV} > a_{IR}$. The $a$-anomaly coefficient can be computed in simple cases like free theories\cite{duff_observations_1977} and weakly coupled theories \cite{opetkou}. Nevertheless we are far from knowing the full set of allowed values of the $a$-anomaly in 4d QFTs.

As shown by Komargodski and Schwimmer \cite{kschwimmer}, the $a$-anomaly of the UV CFT may be probed by coupling a dilaton test particle to the quantum field theory of interest. The dilaton-dilaton scattering amplitude is related to the $a$-anomaly coefficient,
\begin{equation}\label{kstheorem1}
    \mathcal T_{BB\to BB} = \frac{a}{f^4} \left(s^2+t^2+u^2\right) + \dots
\end{equation}
with $a = a_\UV - a_\IR$, or, in case of gapped theories, $a = a_\UV$. In the limit of probe coupling $f\to \infty$, the original theory is unperturbed.
Here we denoted with $B$ the dialaton and $s,t,u$ are Mandelstam invariants, defined by the (all-ingoing) momenta in the process,

\begin{align}
    s &= (p_{1}+p_{2})^2\\
    t &= (p_{1}+p_{3})^2\\
    u &= (p_{1}+p_{4})^2\ .
\end{align}

The space of consistent scattering amplitudes may be investigated from the point of view of the S-matrix bootstrap. For each amplitude, a general ansatz with appropriate pole and branch cut structure is constructed, and then, by finding the range of parameters consistent with unitarity bounds, bounds on quantities of interest (in this case, the $a$-anomaly), can be found. These computations were able to constrain the parameter range in large number of 4d QFT, ranging from $\phi^4$ scalar theories \cite{smatrix3}\cite{smatrix4}\cite{denis_doesnt_spin}, pion scattering in QCDs \cite{lakepenisulakink}\cite{jan_scatters_pions}, or massless vector particles \cite{kelian}

This paper QFT with one or more stable scalar particles, coupled to the dilaton field. The details about setting up the amplitude ansatz with appropriate analytic structure and obeying adequate soft conditions, are given in the next section, along with the  unitarity conditions. This is followed by a section discussing the numerical setup, translating the S-matrix bootstrap task into a semidefinite matrix problem (SDP).

Previously, similar computations \cite{a_anomaly_paper} reproduced the value of $a = \left( 5760\pi^2 \right)^{-1}$ found by analytic CFT computations \cite{opetkou} for the theory of free scalar, and found a lower bound on $a$ for a theory with a single scalar particle and with a $\mathbb{Z}_2$ global symmetry. This follow-up paper generalizes the problem and investigates theories of
\begin{itemize}
    \item single stable scalar particle (no $\mathbb{Z}_2$)
    \item two stable scalar particles (with varying mass ratio)
    \item many stable scalar particles
\end{itemize}

The detailed results of these computations are discussed in detail in section \ref{results}, and the corresponding bounds on the $a$-anomaly are presented in a table below.
\begin{center}
\begin{tabular}{|m{3.5cm}|m{3cm}|m{7cm}|}
    \hline
        Theory &
        $a$-anomaly &
        Annotations \\
    \hline
        Single \textbf{free} scalar &
        $\aFree = \frac{1}{5760\pi^2}$ &
        \small Derived analyticly in \cite{opetkou}, confirm to saturate unitarity bounds \cite{a_anomaly_paper}.\\
    \hline
        Single stable $\mathbb{Z}_2$-odd scalar &
        $\gtrsim 0.32 \cdot \aFree$ &
        \small Investigated in \cite{a_anomaly_paper}.\\
    \hline
        Single stable scalar &
        $\gtrsim 0.15 \cdot \aFree$ &
        \small No resonances at $s < 4m_A^2$, with $m_A$ the mass of the lightest particle.\\
    \hline
        Two stable scalars &
        $\gtrsim 0.034 \cdot \aFree$ &
        \small The minimal $a$-anomaly occurs for $m_X^2 \approx \left(2.5\pm0.1\right)m_A^2$. \\
    \hline
        Many stable scalars &
        $\gtrsim 0.036 \cdot \aFree$ &
        \small    The data comes only from preliminary investigations, which explains why the extrapolated bound is slightly  inconsistent with the previous case.\\
    \hline
\end{tabular}
\end{center}
\section{S-matrix bootstrap setup}\label{setup}

To \textit{probe} a quantum field theory, one introduces the \textbf{dilaton}, a massless scalar particle basically coupled to the trace of the energy-momentum tensor  \cite{kschwimmer}.
The S-matrix bootstrap setup containing the lightest stable scalar particle of the theory, called $A$ (and of mass $m_A$, in practical computations normalized to 1), and dilaton $B$ (of mass $m_B = 0$) contains the following amplitudes:
\begin{equation}
  \begin{gathered}
    AA\rightarrow AA,\quad
    AA\rightarrow AB,\quad
    AA\rightarrow BB,\\
    AB\rightarrow BB,\quad
    BB\rightarrow BB,\quad
    AB\rightarrow AB.
  \end{gathered}
\end{equation}
Note that $AB \to AB$ is $s$-$t$ crossing of $AA\rightarrow BB$, and the other amplitudes are fully crossing symmetric. 

Similarly to case of $\mathbb{Z}_2$-symmetric case described in \cite{a_anomaly_paper}, the nontrivial part of the scattering amplitudes depends on the probe coupling, with 
\begin{equation}\
  \begin{aligned}
    &\mathcal T_{AA\rightarrow AA} = \mathcal {\tilde T}_{AA\rightarrow AA} + O(f^{-1})\\
    &\mathcal T_{AA\rightarrow AB} = \frac 1 f \mathcal {\tilde T}_{AA\rightarrow AB} + O(f^{-2})\\
    &\mathcal T_{AA\rightarrow BB} = \frac 1 {f^2} \mathcal {\tilde T}_{AA\rightarrow BB} + O(f^{-3})\\
    &\mathcal T_{AB\rightarrow BB} = \frac 1 {f^3} \mathcal {\tilde T}_{AB\rightarrow BB} + O(f^{-4})\\
    &\mathcal T_{BB\rightarrow BB} = \frac 1 {f^4} \mathcal {\tilde T}_{BB\rightarrow BB} + O(f^{-5})\\
    &\mathcal T_{AB\rightarrow AB} = \frac 1 {f^2} \mathcal {\tilde T}_{AA\rightarrow BB} + O(f^{-3})
  \end{aligned}
\end{equation}

The amplitude $\mathcal {\tilde T}_{BB\rightarrow BB} = a_\UV\left(s^2+t^2+u^2\right)$, via \cite{kschwimmer}, relates the $a$-anomaly of UV CFT to dilaton scattering and is the  central point of interest of the numerical experiments described in this paper.

\subsection{Unitarity}

To describe the unitarity conditions, it's most feasible to decompose amplitudes (functions $\mathcal T(s,t)$\ ) into partial waves $\mathcal { T}^{\ell}(s)$. The details of such decomposition is described in great extent in \cite{denis_spins_around}. For each value of $s$ above the physical threshold, the unitarity of $2\to 2$ particle scattering can be described using the condition

\begin{equation}
    \begin{pmatrix}
        1 & 0 & 0 & \mathcal { T}^{* \ell}_{AA\rightarrow AA} & \mathcal { T}^{* \ell}_{AA\rightarrow AB} & \mathcal { T}^{* \ell}_{AA\rightarrow BB}\\
        0 & 1 & 0 & \mathcal { T}^{* \ell}_{AB\rightarrow AA} & \mathcal { T}^{* \ell}_{AB\rightarrow AB} & \mathcal { T}^{* \ell}_{AB\rightarrow BB}\\
        0 & 0 & 1 & \mathcal { T}^{* \ell}_{BB\rightarrow AA} & \mathcal { T}^{* \ell}_{BB\rightarrow AB} & \mathcal { T}^{* \ell}_{BB\rightarrow BB}\\
        \mathcal { T}^{\ell}_{AA\rightarrow AA} & \mathcal { T}^{\ell}_{AB\rightarrow AA} & \mathcal { T}^{\ell}_{BB\rightarrow AA}
        & 2\Im\mathcal { T}^{\ell}_{AA\rightarrow AA} & 2\Im\mathcal { T}^{\ell}_{AA\rightarrow AB} & 2\Im\mathcal { T}^{\ell}_{AA\rightarrow BB}\\
        \mathcal { T}^{\ell}_{AA\rightarrow AB} & \mathcal { T}^{\ell}_{AB\rightarrow AB} & \mathcal { T}^{\ell}_{BB\rightarrow AB}
        & 2\Im\mathcal { T}^{\ell}_{AA\rightarrow AB} & 2\Im\mathcal { T}^{\ell}_{AB\rightarrow AB} & 2\Im\mathcal { T}^{\ell}_{AB\rightarrow BB}\\
        \mathcal { T}^{\ell}_{AA\rightarrow BB} & \mathcal { T}^{\ell}_{AB\rightarrow BB} & \mathcal { T}^{\ell}_{BB\rightarrow BB}
        & 2\Im\mathcal { T}^{\ell}_{AA\rightarrow BB} & 2\Im\mathcal { T}^{\ell}_{AB\rightarrow BB} & 2\Im\mathcal { T}^{\ell}_{BB\rightarrow BB} 
    \end{pmatrix} \succcurlyeq 0
\end{equation}
In $f\rightarrow 0$ limit this condition simplifies to

\begin{equation}
    \begin{pmatrix}
        1 & \mathcal {\tilde T}^{* \ell}_{AA\rightarrow AA} & \mathcal {\tilde T}^{* \ell}_{AA\rightarrow AB} & \mathcal {\tilde T}^{* \ell}_{AA\rightarrow BB}\\
        \mathcal {\tilde T}^{\ell}_{AA\rightarrow AA}
        & 2\Im\mathcal {\tilde T}^{\ell}_{AA\rightarrow AA} & 2\Im\mathcal {\tilde T}^{\ell}_{AA\rightarrow AB} & 2\Im\mathcal {\tilde T}^{\ell}_{AA\rightarrow BB}\\
        \mathcal {\tilde T}^{\ell}_{AA\rightarrow AB}
        & 2\Im\mathcal {\tilde T}^{\ell}_{AA\rightarrow AB} & 2\Im\mathcal {\tilde T}^{\ell}_{AB\rightarrow AB} & 2\Im\mathcal {\tilde T}^{\ell}_{AB\rightarrow BB}\\
        \mathcal {\tilde T}^{\ell}_{AA\rightarrow BB}
        & 2\Im\mathcal {\tilde T}^{\ell}_{AA\rightarrow BB} & 2\Im\mathcal {\tilde T}^{\ell}_{AB\rightarrow BB} & 2\Im\mathcal {\tilde T}^{\ell}_{BB\rightarrow BB} 
    \end{pmatrix} \succcurlyeq 0
\end{equation}

Introducing this full matrix ($4\times4$ with 5 independent amplitudes) is computationally very expensive. Instead, via Sylvester's criterion, two necessary conditions are investigated numerically:
\begin{equation}\label{unitarityMatrix}
    \begin{pmatrix}
        1 & \mathcal {\tilde T}^{* \ell}_{AA\rightarrow AA} & \mathcal {\tilde T}^{* \ell}_{AA\rightarrow BB}\\
        \mathcal {\tilde T}^{\ell}_{AA\rightarrow AA}
        & 2\Im\mathcal {\tilde T}^{\ell}_{AA\rightarrow AA} & 2\Im\mathcal {\tilde T}^{\ell}_{AA\rightarrow BB}\\
        \mathcal {\tilde T}^{\ell}_{AA\rightarrow BB}
        & 2\Im\mathcal {\tilde T}^{\ell}_{AA\rightarrow BB} & 2\Im\mathcal {\tilde T}^{\ell}_{BB\rightarrow BB} 
    \end{pmatrix} \succcurlyeq 0, \quad
    \begin{pmatrix} 2\Im\mathcal {\tilde T}^{\ell}_{AB\rightarrow AB}
    \end{pmatrix} \succcurlyeq 0, \quad
\end{equation}
Note that these two conditions are sufficient if processes $AA\rightarrow AB$ and $AB\rightarrow BB$ are forbidden by a $\mathbb{Z}_2$ symmetry of theory as assumed in \cite{a_anomaly_paper}. Looking at this subset of unitarity conditions reduces problem to three independent amplitudes:
  \begin{equation*}
    AA\rightarrow AA,\quad AA\rightarrow BB, \quad BB\rightarrow BB\ .
  \end{equation*} 

\subsection{Analicity and crossing}

The `physical threshold' mentioned previously requires clarification. The usual formula $s > \left(
\mathrm{max}\left(\sum_i m_{i,\textrm{in}}, \sum_i m_{i,\textrm{out}}, \right)
 \right)^2$ would suggest $s>0$. However, in the probe limit $f \rightarrow \infty$, the discontinuities only start as a result of QFT intermediate states, as the dilaton is decoupled from the theory. This implies each amplitude $\tilde {\mathcal T}$ has a branch cut beginning at $s = 4m_A^2$. This provides a natural choice for parametrizing finite part of each amplitude
\begin{align}\label{rhoSeries}
    \mathcal {\tilde T}_{AA\rightarrow AA}(s,t,u) = \sum_{a,b,c} \alpha_{abc} \rhoVar[a]{s}{\frac 43} \rhoVar[b]{t}{\frac 43} \rhoVar[c]{u}{\frac 43} + \dots\\
    \mathcal {\tilde T}_{AA\rightarrow BB}(s,t,u) = \sum_{a,b,c} \beta_{abc} \rhoVar[a]{s}{0} \rhoVar[b]{t}{1} \rhoVar[c]{u}{1} + \dots\\
    \mathcal {\tilde T}_{BB\rightarrow BB}(s,t,u) = \sum_{a,b,c} \gamma_{abc} \rhoVar[a]{s}{0} \rhoVar[b]{t}{0} \rhoVar[c]{u}{0} + \dots
\end{align}
with 
\begin{equation}
    \rhoVar{s}{s_0} = \frac{\sqrt{4m_A^2 - s_0} - \sqrt{4m_A^2 - s}}{\sqrt{4m_A^2 - s_0} + \sqrt{4m_A^2 - s}}
\end{equation}
Indeed, the function $\rhoVar{s}{s_0}$ has the appropriate branch cut starting at $s = 4m_A^2$. The unitarity conditions \eqref{unitarityMatrix} are to be imposed above this cut, at $s\in (4m_A^2; \infty)$. The choice of second parameter of $\rho$ function (the origin of rho series) is to keep crossing relations and soft conditions as simple as possible, which will be justified in next sections.

With amplitudes $AA\rightarrow AA$ and $BB\rightarrow BB$ being fully crossing symmetric, the crossing conditions imply $\alpha_{abc} = \alpha_{bac} = \alpha_{cba}$ and $\gamma_{abc} = \gamma_{bac} = \gamma_{cba}$, and for amplitude $AA\rightarrow BB$, which is symmetric in $tu$ variables, the condition is $\beta_{abc} = \beta_{acb}$, and the amplitude $AB\to AB$ is related by
\begin{equation}
    \mathcal {\tilde T}_{AB\rightarrow AB}(s,t,u) = \mathcal {\tilde T}_{AA\rightarrow BB}(t,s,u)
\end{equation}

The Mandelstam variables $s,t,u$ are not independent, with $s+t+u = \sum m_i^2$, and $m_i$'s being masses of incoming and outgoing particles. To reduce the redundancy in parametrization of $\mathcal {\tilde T}$'s, one imposes $\alpha_{abc} = 0$ if $a\neq0 \land b\neq0 \land c\neq0$ and similar for other amplitudes.

\subsection{Pole structure and soft conditions}

Having a QFT with stable particles of masses $m_A, m_B = 0$ and possibly some other matter content of mass $m_X$ leads to existence of poles in scattering amplitudes.

Keeping in mind crossing symmetries, the poles due to exchange of particle $A$ in aforementioned amplitudes are
\begin{subequations}\label{particleExchangeFullSet}
\begin{equation}
  \begin{aligned}
    \mathcal{\tilde T}_{AA\rightarrow AA} &=
    \begin{tikzpicture}[baseline=($(a.base)!0.5!(b.base)$)]
      \begin{feynman}[small]
        \vertex [blob, centered, pattern=none, node font=\small, inner sep = .3mm] (a) {$g_0$};
        \vertex [right =of a, blob, centered, pattern=none, node font=\small, inner sep = .3mm] (b) {$g_0$};
        \vertex [above left=of a](i1);
        \vertex [above right=of b] (o1) ;
        \vertex [below left=of a] (i2);
        \vertex [below right=of b] (o2);
        \diagram* {
          {(i1), (i2)}--(a)--(b)--{(o1),(o2)}
        };
      \end{feynman}
    \end{tikzpicture}
     +
      \begin{tikzpicture}[baseline=($(a.base)!0.5!(b.base)$)]
        \begin{feynman}[small]
          \vertex [blob, centered, pattern=none, node font=\small, inner sep = .3mm] (a) {$g_0$};
          \vertex [below=of a, blob, centered, pattern=none, node font=\small, inner sep = .3mm] (b) {$g_0$};
          \vertex [above left=of a](i1);
          \vertex [above right=of a] (o1) ;
          \vertex [below left=of b] (i2);
          \vertex [below right=of b] (o2);
          \diagram* {
            {(i1),(o1)}--(a)--(b)--{(i2),(o2)}
          };
        \end{feynman}
      \end{tikzpicture}
      +
      \begin{tikzpicture}[baseline=($(a.base)!0.5!(b.base)$)]
        \begin{feynman}[small]
          \vertex [blob, centered, pattern=none, node font=\small, inner sep = .3mm] (a) {$g_0$};
          \vertex [below=of a, blob, centered, pattern=none, node font=\small, inner sep = .3mm] (b) {$g_0$};
          \vertex [above left=of a](i1);
          \vertex [above right=of a] (o1) ;
          \vertex [below left=of b] (i2);
          \vertex [below right=of b] (o2);
          \diagram* {
            {(i1),(o2)}--(a)--(b)--{(i2),(o1)}
          };
        \end{feynman}
      \end{tikzpicture}
      + \dots =\\
      &= -\left|g_0\right|^2 \left(\frac{1}{s-m^2_A}+\frac{1}{t-m^2_A}+\frac{1}{u-m^2_A}\right) + \dots
  \end{aligned}
\end{equation}

\begin{equation}
  \begin{aligned}
    \mathcal{\tilde T}_{AA\rightarrow BB} &=
    \begin{tikzpicture}[baseline=($(a.base)!0.5!(b.base)$)]
      \begin{feynman}[small]
        \vertex [blob, centered, pattern=none, node font=\small, inner sep = .3mm] (a) {$g_0$};
        \vertex [right =of a, blob, centered, pattern=none, node font=\small, inner sep = .3mm] (b) {$g_2$};
        \vertex [above left=of a](i1);
        \vertex [above right=of b] (o1) ;
        \vertex [below left=of a] (i2);
        \vertex [below right=of b] (o2);
        \diagram* {
          {(i1), (i2)}--(a)--(b)--[dashed]{(o1),(o2)}
        };
      \end{feynman}
    \end{tikzpicture}
     +
      \begin{tikzpicture}[baseline=($(a.base)!0.5!(b.base)$)]
        \begin{feynman}[small]
          \vertex [blob, centered, pattern=none, node font=\small, inner sep = .3mm] (a) {$g_1$};
          \vertex [below=of a, blob, centered, pattern=none, node font=\small, inner sep = .3mm] (b) {$g_1$};
          \vertex [above left=of a](i1);
          \vertex [above right=of a] (o1) ;
          \vertex [below left=of b] (i2);
          \vertex [below right=of b] (o2);
          \diagram* {
            (i1)--(a)--(b)--(i2),
            (o1)--[dashed](a),
            (o2)--[dashed](b),
          };
        \end{feynman}
      \end{tikzpicture}
      +
      \begin{tikzpicture}[baseline=($(a.base)!0.5!(b.base)$)]
        \begin{feynman}[small]
          \vertex [blob, centered, pattern=none, node font=\small, inner sep = .3mm] (a) {$g_1$};
          \vertex [below=of a, blob, centered, pattern=none, node font=\small, inner sep = .3mm] (b) {$g_1$};
          \vertex [above left=of a](i1);
          \vertex [above right=of a] (o1) ;
          \vertex [below left=of b] (i2);
          \vertex [below right=of b] (o2);
          \diagram* {
            (i1)--(a)--(b)--(i2),
            (o2)--[dashed](a),
            (o1)--[dashed](b),
          };
        \end{feynman}
      \end{tikzpicture}+ \dots = \\
      &= -\frac{g_0 g_2}{s-m^2_A} - \left|g_1\right|^2\left(\frac{1}{t-m^2_A}+\frac{1}{u-m^2_A}\right) + \dots
\end{aligned}
\end{equation}

\begin{equation}
  \begin{aligned}
    \mathcal{\tilde T}_{BB\rightarrow BB} &=
    \begin{tikzpicture}[baseline=($(a.base)!0.5!(b.base)$)]
      \begin{feynman}[small]
        \vertex [blob, centered, pattern=none, node font=\small, inner sep = .3mm] (a) {$g_2$};
        \vertex [right =of a, blob, centered, pattern=none, node font=\small, inner sep = .3mm] (b) {$g_2$};
        \vertex [above left=of a](i1);
        \vertex [above right=of b] (o1) ;
        \vertex [below left=of a] (i2);
        \vertex [below right=of b] (o2);
        \diagram* {
          {(i1), (i2)}--[dashed](a)--(b)--[dashed]{(o1),(o2)}
        };
      \end{feynman}
    \end{tikzpicture}
     +
      \begin{tikzpicture}[baseline=($(a.base)!0.5!(b.base)$)]
        \begin{feynman}[small]
          \vertex [blob, centered, pattern=none, node font=\small, inner sep = .3mm] (a) {$g_2$};
          \vertex [below=of a, blob, centered, pattern=none, node font=\small, inner sep = .3mm] (b) {$g_2$};
          \vertex [above left=of a](i1);
          \vertex [above right=of a] (o1) ;
          \vertex [below left=of b] (i2);
          \vertex [below right=of b] (o2);
          \diagram* {
            {(i1),(o1)}--[dashed](a)--(b)--[dashed]{(i2),(o2)}
          };
        \end{feynman}
      \end{tikzpicture}
      +
      \begin{tikzpicture}[baseline=($(a.base)!0.5!(b.base)$)]
        \begin{feynman}[small]
          \vertex [blob, centered, pattern=none, node font=\small, inner sep = .3mm] (a) {$g_2$};
          \vertex [below=of a, blob, centered, pattern=none, node font=\small, inner sep = .3mm] (b) {$g_2$};
          \vertex [above left=of a](i1);
          \vertex [above right=of a] (o1) ;
          \vertex [below left=of b] (i2);
          \vertex [below right=of b] (o2);
          \diagram* {
            {(i1),(o2)}--[dashed](a)--(b)--[dashed]{(i2),(o1)}
          };
        \end{feynman}
      \end{tikzpicture}+ \dots\\
      &= -\left|g_2\right|^2 \left(\frac{1}{s-m^2_A}+\frac{1}{t-m^2_A}+\frac{1}{u-m^2_A}\right) + \dots
  \end{aligned}
\end{equation}
\end{subequations}
where solid line is propagation of $A$ and dashed line is propagation of $B$.
If one introduces an additional particle $m_X$, other poles are introduced due to its exchange:
\begin{subequations}\label{xExchangeFullSet}
\begin{equation}
  \begin{aligned}
  \mathcal{\tilde T}_{AA\rightarrow AA} &=
  \begin{tikzpicture}[baseline=($(a.base)!0.5!(b.base)$)]
    \begin{feynman}[small]
      \vertex [blob, centered, pattern=none, node font=\small, inner sep = .3mm] (a) {$g_0'$};
      \vertex [right =of a, blob, centered, pattern=none, node font=\small, inner sep = .3mm] (b) {$g_0'$};
      \vertex [above left=of a](i1);
      \vertex [above right=of b] (o1) ;
      \vertex [below left=of a] (i2);
      \vertex [below right=of b] (o2);
      \diagram* {
        {(i1), (i2)}--(a)--[double](b)--{(o1),(o2)}
      };
    \end{feynman}
  \end{tikzpicture}
   +
    \begin{tikzpicture}[baseline=($(a.base)!0.5!(b.base)$)]
      \begin{feynman}[small]
        \vertex [blob, centered, pattern=none, node font=\small, inner sep = .3mm] (a) {$g_0'$};
        \vertex [below=of a, blob, centered, pattern=none, node font=\small, inner sep = .3mm] (b) {$g_0'$};
        \vertex [above left=of a](i1);
        \vertex [above right=of a] (o1) ;
        \vertex [below left=of b] (i2);
        \vertex [below right=of b] (o2);
        \diagram* {
          {(i1),(o1)}--(a)--[double](b)--{(i2),(o2)}
        };
      \end{feynman}
    \end{tikzpicture}
    +
    \begin{tikzpicture}[baseline=($(a.base)!0.5!(b.base)$)]
      \begin{feynman}[small]
        \vertex [blob, centered, pattern=none, node font=\small, inner sep = .3mm] (a) {$g_0'$};
        \vertex [below=of a, blob, centered, pattern=none, node font=\small, inner sep = .3mm] (b) {$g_0'$};
        \vertex [above left=of a](i1);
        \vertex [above right=of a] (o1) ;
        \vertex [below left=of b] (i2);
        \vertex [below right=of b] (o2);
        \diagram* {
          {(i1),(o2)}--(a)--[double](b)--{(i2),(o1)}
        };
      \end{feynman}
    \end{tikzpicture} \dots =\\
    &= -\left|g'_0\right|^2 \left(\frac{1}{s-m^2_X}+\frac{1}{t-m^2_X}+\frac{1}{u-m^2_X}\right) + \dots
  \end{aligned}
\end{equation}

\begin{align}
  \mathcal{\tilde T}_{AA\rightarrow BB} &=
    \begin{tikzpicture}[baseline=($(a.base)!0.5!(b.base)$)]
      \begin{feynman}[small]
        \vertex [blob, centered, pattern=none, node font=\small, inner sep = .3mm] (a) {$g_0'$};
        \vertex [right =of a, blob, centered, pattern=none, node font=\small, inner sep = .3mm] (b) {$g_2'$};
        \vertex [above left=of a](i1);
        \vertex [above right=of b] (o1) ;
        \vertex [below left=of a] (i2);
        \vertex [below right=of b] (o2);
        \diagram* {
          {(i1), (i2)}--(a)--[double](b)--[dashed]{(o1),(o2)}
        };
      \end{feynman}
    \end{tikzpicture} + \dots = -\frac{g'_0 g'_2}{s-m^2_X} + \dots
\end{align}

\begin{equation}\label{xParticleTbb}
  \begin{aligned}
  \mathcal{\tilde T}_{BB\rightarrow BB} &=
  \begin{tikzpicture}[baseline=($(a.base)!0.5!(b.base)$)]
    \begin{feynman}[small]
      \vertex [blob, centered, pattern=none, node font=\small, inner sep = .3mm] (a) {$g_2'$};
      \vertex [right =of a, blob, centered, pattern=none, node font=\small, inner sep = .3mm] (b) {$g_2'$};
      \vertex [above left=of a](i1);
      \vertex [above right=of b] (o1) ;
      \vertex [below left=of a] (i2);
      \vertex [below right=of b] (o2);
      \diagram* {
        {(i1), (i2)}--[dashed](a)--[double](b)--[dashed]{(o1),(o2)}
      };
    \end{feynman}
  \end{tikzpicture}
   +
    \begin{tikzpicture}[baseline=($(a.base)!0.5!(b.base)$)]
      \begin{feynman}[small]
        \vertex [blob, centered, pattern=none, node font=\small, inner sep = .3mm] (a) {$g_2'$};
        \vertex [below=of a, blob, centered, pattern=none, node font=\small, inner sep = .3mm] (b) {$g_2'$};
        \vertex [above left=of a](i1);
        \vertex [above right=of a] (o1) ;
        \vertex [below left=of b] (i2);
        \vertex [below right=of b] (o2);
        \diagram* {
          {(i1),(o1)}--[dashed](a)--[double](b)--[dashed]{(i2),(o2)}
        };
      \end{feynman}
    \end{tikzpicture}
    +
    \begin{tikzpicture}[baseline=($(a.base)!0.5!(b.base)$)]
      \begin{feynman}[small]
        \vertex [blob, centered, pattern=none, node font=\small, inner sep = .3mm] (a) {$g_2'$};
        \vertex [below=of a, blob, centered, pattern=none, node font=\small, inner sep = .3mm] (b) {$g_2'$};
        \vertex [above left=of a](i1);
        \vertex [above right=of a] (o1) ;
        \vertex [below left=of b] (i2);
        \vertex [below right=of b] (o2);
        \diagram* {
          {(i1),(o2)}--[dashed](a)--[double](b)--[dashed]{(i2),(o1)}
        };
      \end{feynman}
    \end{tikzpicture} + \dots =\\
    &= -\left|g_2'\right|^2 \left(\frac{1}{s-m^2_X}+\frac{1}{t-m^2_X}+\frac{1}{u-m^2_X}\right) + \dots
  \end{aligned}
\end{equation}
The vertex factor
\begin{equation}
  \begin{tikzpicture}[baseline=($(a.base)$)]
    \begin{feynman}[small]
      \vertex [blob, centered, pattern=none, node font=\small, inner sep = .3mm] (a) {$g_1'$};
      \vertex [left=of a](i1);
      \vertex [above right=of a](i2);
      \vertex [below right=of a](i3);
      \diagram* {
        (i1) -- (a);
        (i2) --[double] (a);
        (i3) --[dashed] (a)
      };
    \end{feynman}
  \end{tikzpicture} = 0
\end{equation}
\end{subequations}
This is a result of construction of dilaton effective action. As detailed in \cite{a_anomaly_paper}, the dilaton couples to relevant operators (usually masses) in perturbed CFT action. With only diagonalized mass terms (like $m_A^2 \phi_A^2$ and $m_X^2 \phi_X^2$) no effective term including dilaton and two different types of matter may arise. 

Same construction also fixes some residues of $\mathcal {\tilde T}_{AA\rightarrow BB}$, as well as the finite part of amplitude in their vicinity. The construction, described in more details in \cite{a_anomaly_paper} gives
\begin{align}\label{softCondition}
\mathcal{\tilde T}_{AA\rightarrow BB} = - 2 \left(\frac{m_A^4}{t-m_A^2} + \frac{m_A^4}{u-m_A^2} \right) - m_A^2 + O(s,t-m_A^2, u-m_A^2)
\end{align}
around the point $s=0,t=m_A^2,u=m_A^2$. 

The final soft condition, given  by \eqref{kstheorem1}, also defines a goal of bootstrap experiment.
\begin{align}\label{kgTheorem}
  \mathcal{\tilde T}_{BB\rightarrow BB} = a\left(s^2+t^2+u^2\right) + \dots
\end{align}

The unitarity conditions on $\mathcal{\tilde T}_{BB\rightarrow BB}$ will provide a non-zero bound on value of $a$-anomaly in theories containing spin-0 particles. The next section provides details about writing these conditions in the form of a semidefinite problem, which will then be investigated numerically.
\section{Numerical implementation}\label{numerical}

In practical computations, the amplitude is parametrized as linear combination of `building blocks':
\begin{align}
    \mathcal T(s,t,u) = \sum \mathtt{coeff}_i \cdot f_i(s,t,u)
\end{align}

Constructing an ansatz with analytic properties described in previous section, and imposing unitarity conditions on a set of matrices \eqref{unitarityMatrix} derived from these amplitudes\footnote{The matrices are therefore linear functions of $\mathtt{coeff}_i$ as well.} allows to form the question about $a$-anomaly bounds as a semidefinite matrix problem (SDP), which allows using specialized solvers.

Therefore, using SDPB\cite{sdpb} to find a vector $\mathtt{coeff}_i$ that minimizes $a$-anomaly given by \eqref{kgTheorem}, and describes amplitudes that obey unitarity bounds \eqref{unitarityMatrix} can be done.
However, besides including $\rho$ series, described by \eqref{rhoSeries} (already a linear combination), the ansatz need to contain poles, and follow soft conditions described in previous section.

\subsection{Poles}

As mentioned previously, the terms in amplitude resulting from exchange of \eqref{particleExchangeFullSet} are non-linear functions of coupling constants $g_0, g_1, g_2$. To be able to put the problem into semidefinite linear form, one shall define `independent' linear coefficients
\begin{equation}
    \begin{aligned}
    \texttt{ga} &\coloneq g_0^2\\
    \texttt{gb} &\coloneq g_1^2\\
    \texttt{gbb} &\coloneq g_0 g_2\\
    \texttt{gdila} &\coloneq g_2^2
    \end{aligned}
\end{equation}

For this parametrization to be equivalent, these parameters have to obey equation
\begin{equation}\label{polesRelation}
    \texttt{ga}\cdot\texttt{gdila} = \texttt{gbb}\cdot \texttt{gbb}
\end{equation}
which may be written as a condition on matrix determinant

\begin{align}
    \begin{vmatrix}
        \texttt{ga} &\texttt{gbb}\\
        \texttt{gbb} &\texttt{gdila}
    \end{vmatrix} = 0
\end{align}

As $\texttt{ga} \ge 0$ and $\texttt{gdila} \ge 0$, the semidefinitness of such matrix

\begin{align}\label{semidefiniteTrick}
    \begin{pmatrix}
        \texttt{ga} &\texttt{gbb}\\
        \texttt{gbb} &\texttt{gdila}
    \end{pmatrix} \succeq 0
\end{align}
does impose the inequality
\begin{equation}\label{semidefiniteTrickIneq}
    \texttt{ga}\cdot\texttt{gdila} \ge \texttt{gbb}\cdot \texttt{gbb}
\end{equation}
without any additional unwanted conditions.

As the term in amplitude $\mathcal{\tilde{T}}_{BB\to BB}$ related to parameter $\texttt{gdila}$, to be precise
\begin{equation}
    \mathcal {\tilde T}_{BB\rightarrow BB} \supset -\texttt{gdila}\cdot\left(\frac{1}{s-1}+\frac{1}{t-1}+\frac{1}{u-1}\right)\quad,
\end{equation}
is purely real, and the unitarity conditions \eqref{unitarityMatrix} contain only $\Im \mathcal{\tilde{T}}_{BB\to BB}$, there is no additional bound on \texttt{gdila} than imposed by \eqref{semidefiniteTrick}. In later section it will be shown that to minimize $a$-anomaly \texttt{gdila} is also to be minimized, and without any further constraints, that will lead to \eqref{semidefiniteTrick} being saturated, and \eqref{polesRelation} holding true.

\subsection{Soft conditions}

To impose soft conditions \eqref{softCondition} around points $s=0, t=0, u=0$, first, one fixes $\mathtt{gb} = 2$ to match the residue, and then changes ansatz terms related to $\mathtt{gbb}$. With $\rhoVar{t}{1} = \mathcal O(t-1)$, and $\rhoVar{u}{1} = \mathcal O(u-1)$ this is particularly simple:

\begin{align}
    \mathcal {\tilde T}_{AA\rightarrow BB} =&
        - 2\left(\frac{1}{t-1} + \frac{1}{u-1}\right) -\texttt{gbb}\left(1 + \frac{1}{s-1}\right)+\\
        &+\sum_{a,b,c} \beta_{abc} \rhoVar{s}{0}^a\rhoVar{t}{1}^b\rhoVar{u}{1}^c
\end{align}
That guarantees the finite part of amplitude to be independent of value of $\mathtt{gbb}$, therefore, all needed to impose correct soft condition is setting $\beta_{000}=-1$.

\subsection{The goal: \texorpdfstring{$a$-anomaly}{a-anomaly}}

To relate $a$-anomaly to coefficients of ansatz one needs to compare the relation
\begin{align}
    \mathcal {\tilde T}_{BB\rightarrow BB} = a\left(s^2+t^2+u^2\right) + \dots
\end{align}
to terms of the ansatz. Expanding the ansatz
\begin{align}
    \mathcal {\tilde T}_{BB\rightarrow BB} =
        &-\mathtt{gdila} \left(\frac{1}{s-1}+\frac{1}{t-1}+\frac{1}{u-1}\right)+\\
        &+\sum_{a,b,c}\gamma_{abc}\ \rhoVar{s}{0}^a\rhoVar{t}{0}^b\rhoVar{u}{0}^c
\end{align}
around $s=t=u=0$ gives
\begin{align}
    \mathcal {\tilde T}_{BB\rightarrow BB} = \textrm{(const.)} + \left(\mathtt{gdila} + \frac{\gamma_{001}}{128} + \frac{\gamma_{002}}{256} - \frac{\gamma_{011}}{512}\right)\left(s^2+t^2+u^2\right) + \dots
\end{align}

The constant term is purely real, so, in practice, as the unitarity matrices contain only $\Im \mathcal{\tilde T}_{BB\to BB}$, doesn't have to be explicitly fixed to 0.  

\subsection{Additional particles}

The procedure of including another particles, as described by \eqref{xExchangeFullSet}, follows very similarly. Again, one has to introduce `independent' linear coefficients

\begin{equation}
    \begin{aligned}
    \texttt{gaX} &\coloneq g_0^2\\
    \texttt{gbbX} &\coloneq g_0 g_2\\
    \texttt{gdilaX} &\coloneq g_2^2
    \end{aligned}
\end{equation}

with semidefinite condition

\begin{align}
    \begin{pmatrix}\label{semidefiniteTrickIneqX}
        \texttt{gaX} &\texttt{gbbX}\\
        \texttt{gbbX} &\texttt{gdilaX}
    \end{pmatrix} \succeq 0\quad .
\end{align}

To keep the soft conditions on $\mathcal{\tilde T}_{AA\to BB}$ intact, the amplitude term related to $\texttt{gbbX}$ is
\begin{align}
    \mathcal {\tilde T}_{AA\rightarrow BB} \supset -\texttt{gbb}\left(\frac{1}{m_X^2} + \frac{1}{s-m_X^2}\right)
\end{align} 
which again, has no contribution to finite part of amplitude around $s=0, t=u=1$.

The contribution to $a$-anomaly from terms described in \eqref{xParticleTbb}, is derived as before, giving additional contribution of

\begin{equation}
    \mathcal {\tilde T}_{BB\rightarrow BB} = \dots + \frac{\mathtt{gdilaX}}{m_X^6}\left(s^2+t^2+u^2\right) + \dots
\end{equation}

Such construction can be repeated to include arbitrary many resonances resulting from exchange of particles of masses $1<m_X<2$. The condition \eqref{semidefiniteTrickIneqX} will again always be saturated when minimizing $a$-anomaly.
\subsection{Improvement terms}
Although, any amplitude of analytical properties described before can be reproduced by a linear combination of resonances, and (infinite) $\rho$ series mentioned before, in practical computations one needs to limit the experiment to finite number of terms, usually with $\alpha_{abc} \neq 0$ only for some $a+b+c \leq \maxN$. Given these limitations, it is often beneficial to include `redundant' terms in the ansatz to improve convergence in such case.

The `threshold singularity term', as introduced in \cite{smatrix3} corresponds to bound state of two $A$ particles, and is included in $AA\to AA$ amplitude ansatz as

\begin{equation}
    \mathcal {\tilde T}_{AA\rightarrow AA}
    \supset
    \xi \cdot \left(
        \frac{1}{\rhoVar{s}{\frac 4 3}-1} + 
        \frac{1}{\rhoVar{t}{\frac 4 3}-1} + 
        \frac{1}{\rhoVar{u}{\frac 4 3}-1}
    \right)
\end{equation}
with analytic bound on the related coefficient $\xi \in \left[\xi_\textrm{min}, 0\right]$, with $\xi_\textrm{min} = - 32 \sqrt 6 \pi$. The early trials (not plotted in this paper) showed better convergence with such addition, and its behavior in experimental data is discussed in the next section.

Along with improvement terms in amplitude $AA \to AA$, as described in greater detail in \cite{a_anomaly_paper}, dilaton-to-dilaton scattering scattering amplitude can be expanded by including term proportional to $\mathcal{\tilde T}_{BB\to BB}^{\textrm{free}}$, corresponding amplitude in theory of free massive boson (so derived from $\mathcal{\tilde T}_{AA\to AA} = 0$). Details of such derivation were described extensively in \cite{a_anomaly_paper}, and, quoting the relevant part,
\begin{equation}
    \begin{aligned}
    \Im[\widetilde{\mathcal{T}}_{BB\rightarrow BB}^\textrm{free}(s,t)]&=\ \frac{1}{32\pi}\sqrt{1-\frac{4}{s}} -\frac{1}{4\pi s}\ln\Bigg(\frac{1+\sqrt{1-\frac{4}{s}}}{1-\sqrt{1-\frac{4}{s}}}\Bigg)\\
    &-\frac{1}{4\pi}\ \frac{1}{su}\frac{1}{\sqrt{1+\frac{4t}{su}}}\ln\Bigg(\frac{\frac{1}{s}-\frac{u}{st}+\frac{u}{2t}\Big[1+\sqrt{1-\frac{4}{s}}\sqrt{1+\frac{4t}{us}}\Big]}{\frac{1}{s}-\frac{u}{st}+\frac{u}{2t}\Big[1-\sqrt{1-\frac{4}{s}}\sqrt{1+\frac{4t}{us}}\Big]}\Bigg)\\
    &-\frac{1}{4\pi}\ \frac{1}{st}\frac{1}{\sqrt{1+\frac{4u}{st}}}\ln\Bigg(\frac{\frac{1}{s}-\frac{t}{su}+\frac{t}{2u}\Big[1+\sqrt{1-\frac{4}{s}}\sqrt{1+\frac{4u}{ts}}\Big]}{\frac{1}{s}-\frac{t}{su}+\frac{t}{2u}\Big[1-\sqrt{1-\frac{4}{s}}\sqrt{1+\frac{4u}{ts}}\Big]}\Bigg)
    \end{aligned}
\end{equation}
and 
\begin{align}
    \mathcal {\tilde T}^{\textrm{free}}_{BB\rightarrow BB} = \aFree\left(s^2+t^2+u^2\right) + \dots
\end{align}
with $\aFree = \frac{1}{5760\pi^2}$.

Expanding ansatz with a term
\begin{equation}
    \mathcal {\tilde T}_{BB\rightarrow BB}
    \supset \mathtt{freeAmp} \cdot \frac{\mathcal{\tilde T}_{BB\rightarrow BB}}{\aFree}
\end{equation}
along with other contribution gives a goal for the optimization problem
\begin{align}
    a = \mathtt{freeAmp} + \mathtt{gdila} + \underbrace{\frac{\mathtt{gdilaX}}{m_X^6}}_{\mathclap{\textrm{or sum of such terms for multiple resonances}}}+ \frac{\gamma_{001}}{128} + \frac{\gamma_{002}}{256} - \frac{\gamma_{011}}{512}
\end{align}

\subsection{The grid, the limitations, the (CPU) time.}

The unitarity condition \eqref{unitarityMatrix} has to be imposed on every $s \in \left(4, \infty\right)$. In practice, SDP computations are limited to finite number of samples in $s$. The grid of values of $s$ has to be chosen carefully to ensure convergence. As the most variation in $\mathcal T$'s occur around physical threshold $s \gtrsim 4$\footnote{This statement is intentionally broad, and comes from trials and errors. If an answer to SDP was computed on too small grid, the unitarity violations between grid points are usually found in vicinity of threshold.}, and shall cover entire range from $4$ to infinity somewhat uniformly (on log scale) for large energies.

The grid used for the computations is based on Chebyshev grid in $\rho$ variables. With
\begin{align*}
    &\rhoVar{4}{0} = 1\\
    &\rhoVar{\infty}{0} = -1
\end{align*}
we introduced grid of
\begin{equation*}
    \delta_k = \frac{1 + \cos \left(\frac{2k-1}{2n}\pi\right)}{2}
\end{equation*}
for $n$ grid points (and $\delta_i \in (0,1)$), which is used to construct a grid of values in $s$ via relation
\begin{equation*}
    \rhoVar{s_k}{0} = e^{i \pi \delta_k}\ .
\end{equation*}
The resulting grid of points $s_i$ has the desired properties mentioned before (large number of points around threshold, and rapidly increasing spacing between points for large $s$, allowing to \textit{decently probe the infinity}).

With pilot computations to establish sufficient grid size, we found no difference between computations made for $n=250$ and $n=300$ grid points in $s$, and smaller sizes, like $n=100$, $n=150$ giving different, and explicitly wrong answers. All computations presented in following section were made using grid of $300$ values in $s$, constructed using algorithm above.

The main building block of each amplitude is the $\rho$ series, 
\begin{align}
    \mathcal {\tilde T} = \sum_{a,b,c} \alpha_{abc} \rhoVar{s}{s_0}^a\rhoVar{t}{t_0}^b\rhoVar{u}{u_0}^c + \cdots
\end{align}
that must be terminated for real computations. With time complexity of problem growing like $O\left(\left|\texttt{coeff}_i\right|^3\right)$ \cite{sdpb}, the evaluation time sharply grows with number of terms in $\rho$ series. Truncating each $\rho$ series by imposing $\alpha_{abc} = 0$ for $a+b+c > \maxN$ is how limiting the number of free coefficients is done. However, with overall time complexity $O\!\left(\maxN^6\right)$, the line between \textit{almost impossible} and \textit{fast and inexpensive} SDPB computation is thin. This line, with hardware accessible for computations in this paper is around $\maxN=40$. The convergence with $\maxN$ is something that needs to be discussed separately for each experiment.

Other limitation is the number of spins included in \eqref{unitarityMatrix}. The computation time grows linearly with number of spins included, so pushing $\maxJ$ to the safe side is not computationally prohibitive. The experiment in this paper show that $\maxJ = \maxN + 12$ is a safe choice, and it is used for later numerics.

The precision of numerical values is another important topic in bootstrap computations. Many term cancelations in computations need large precision floating point numbers, and the experiments described later are no exception. 512 bits of mantissa precision was found to be (safely) more than enough and was used in the numerics for this paper.
\newpage
\section{Results}\label{results}

\subsection{A story of one particle}

\begin{figure}[H]
    \begin{tikzpicture}
    \begin{axis}[
        width=14cm,
        height=8cm,
        xlabel={$\maxN^{-1}$},
        ylabel={$a/ \aFree$},
        legend cell align=left,
        legend style={
            at={(1.02,0.5)},
            anchor=west,nodes={inner sep=0.1cm,text depth=0.0em},
            },
        ymajorgrids=true,
        grid style=dashed,
        scatter/classes={
            0={mark=*, plot1},
            4={mark=diamond*, plot2},
            8={mark=square*, plot3},
            12={mark=triangle*, plot4}
        },
        xmin=0,
        xmax=0.055,
        ymin=0.125,
        ymax=0.25,
        scaled ticks=false,
        xticklabel style={
        /pgf/number format/fixed}
    ]

    \addlegendimage{empty legend}
    \addplot[
        only marks,
        scatter,
        mark size=1.5pt,scatter src=explicit]
    table[
        x="oneOverN",
        y="aAnomaly",
        meta="nMinusJ",
        col sep=comma]
    {data/singlePoleData.csv};
    \addplot[domain=0:0.055, dashed]{0.15117 + 1.26695*x};
    \legend{{\hspace{-0.6cm}\maxJ-\maxN:}, 0, 4, 8, 12, Best fit}
\end{axis}
\end{tikzpicture}\vspace{-0.8cm}
    \centering
    \caption{Absolute minimum of $a$ as a fraction of $\aFree$. A line fit to the best available data (highest \maxJ) is marked.}
    \label{singleconvergence}
\end{figure}
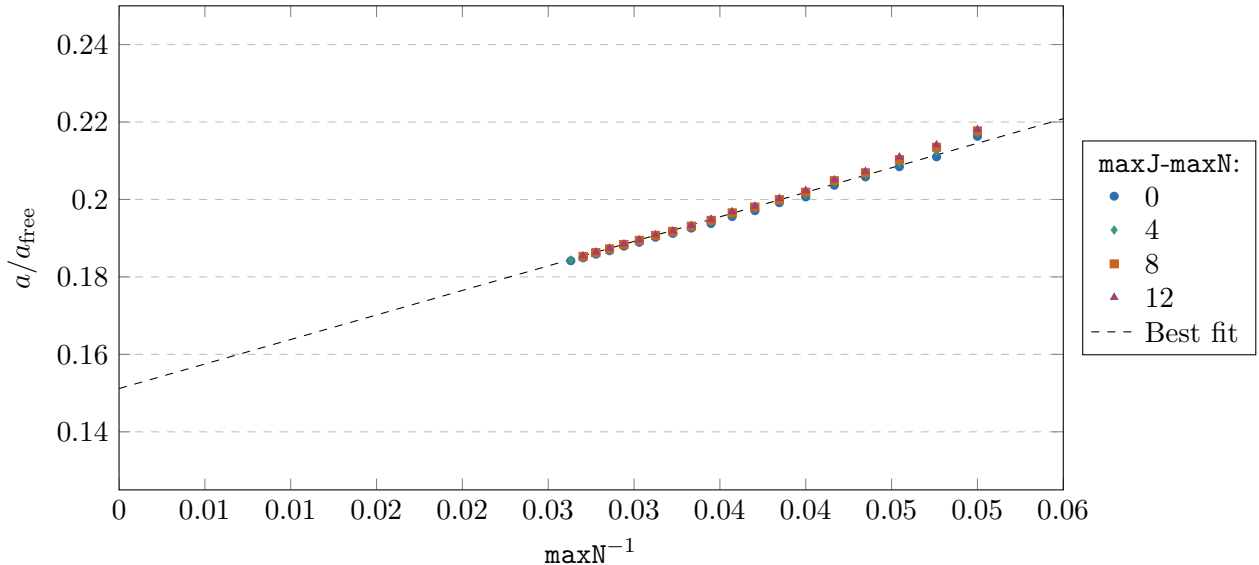

The first performed experiment includes matter (particle $A$) and dilaton (particle $B$) without additional residues. The numerical data allowed to find a feasible range of $\rho$ series size \maxN and number of partial waves \maxJ included in unitarity constraints, and the convergence pattern is shown on figure \ref{singleconvergence}.

The close-to-linear behavior of $\frac{a}{\aFree}$ as a function of $\frac{1}{\maxN}$ allows to extrapolate the limit of $\maxN \rightarrow \infty$, giving the answer of $a \gtrsim 0.1517\cdot\aFree$. This can be concluded to be absolute minimum $a$-anomaly of any theory with a single particle of spin 0.

As presented on a figure, the number of spins required for convergence depends on the ansatz size, however over entire range of feasible \maxN (as time complexity grows approximately as $O(\maxN^6)$ and experiments with $\maxN > 40$ became prohibitively expensive), the difference between solutions found with $\maxJ = \maxN+12$ and $\maxJ = \maxN+8$ differed only marginally, proving $\maxJ = \maxN+12$ is a safe choice for next experiments.

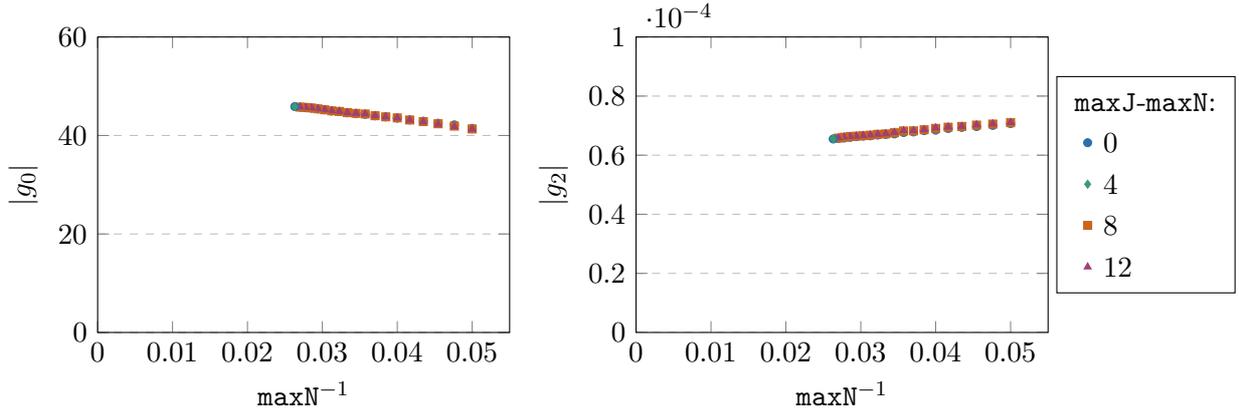
\begin{figure}[H]
    \hspace{-2.5cm}\begin{tikzpicture}
    \begin{axis}[
        width=7cm,
        height=5.5cm,
        xlabel={$\maxN^{-1}$},
        ylabel={$\left|g_0\right|$},
        ymajorgrids=true,
        grid style=dashed,
        scatter/classes={
            0={mark=*, plot1},
            4={mark=diamond*, plot2},
            8={mark=square*, plot3},
            12={mark=triangle*, plot4}
        },
        xmin=0,
        xmax=0.055,
        ymin=0,
        ymax=60,
        scaled ticks=false,
        xticklabel style={
        /pgf/number format/fixed}
    ]
    \addlegendimage{empty legend}
    \addplot[
        only marks,
        scatter,
        mark size=1.5pt,scatter src=explicit]
    table[
        x="oneOverN",
        y="g0",
        meta="nMinusJ",
        col sep=comma]
    {data/singlePoleData.csv};
\end{axis}
\end{tikzpicture}%
\hspace{0.25cm}%
    \begin{tikzpicture}
    \begin{axis}[
        width=7cm,
        height=5.5cm,
        legend cell align=left,
        legend style={
            at={(1.02,0.5)},
            anchor=west,nodes={inner sep=0.15cm,text depth=0.0em},
            },
        xlabel={$\maxN^{-1}$},
        ylabel={$\left|g_2\right|$},
        ymajorgrids=true,
        grid style=dashed,
        scatter/classes={
            0={mark=*, plot1},
            4={mark=diamond*, plot2},
            8={mark=square*, plot3},
            12={mark=triangle*, plot4}
        },
        xmin=0,
        xmax=0.055,
        ymin=0,
        ymax=0.00010,
        scaled x ticks=false,
        xticklabel style={
        /pgf/number format/fixed}
    ]
    \addlegendimage{empty legend}
    \addplot[
        only marks,
        scatter,
        mark size=1.5pt,scatter src=explicit]
    table[
        x="oneOverN",
        y="g2",
        meta="nMinusJ",
        col sep=comma]
    {data/singlePoleData.csv};
    \legend{{\hspace{-0.5cm} \maxJ-\maxN:}, 0, 4, 8, 12}
\end{axis}
\end{tikzpicture}%
\hspace{-2cm}
    \centering
    \caption{Size of 3-point couplings $g_0$ and $g_2$. Note difference in scales.}
    \label{singlecouplings}
\end{figure}

The strength of 3-point couplings may be extracted from data. The residue of pole in $\mathcal{\tilde T}_{AA\rightarrow AA}$ saturates the previously found bounds \cite{smatrix3}, however pole in $\mathcal{\tilde T}_{BB\rightarrow BB}$ does not contribute significantly to $a$-anomaly (see figure \ref{singlecouplings}).

\begin{figure}[ht]
    \hspace{-2.5cm}\begin{tikzpicture}
    \begin{axis}[
        width=7cm,
        height=5.5cm,
        xlabel={$\maxN^{-1}$},
        ylabel={$\xi/ \xi_{\textrm{min}}$},,ylabel shift = -0.5em,
        ymajorgrids=true,
        grid style=dashed,
        scatter/classes={
            0={mark=*, plot1},
            4={mark=diamond*, plot2},
            8={mark=square*, plot3},
            12={mark=triangle*, plot4}
        },
        xmin=0,
        xmax=0.055,
        ymin=0,
        ymax=0.2,
        scaled x ticks=false,
        xticklabel style={
        /pgf/number format/fixed},
        yticklabel style={
        /pgf/number format/fixed},
    ]
    \addlegendimage{empty legend}
    \addplot[
        only marks,
        scatter,
        mark size=1.5pt,scatter src=explicit]
    table[
        x="oneOverN",
        y="xi",
        meta="nMinusJ",
        col sep=comma]
    {data/singlePoleData.csv};
\end{axis}
\end{tikzpicture}\hspace{0.25cm}%
    \begin{tikzpicture}
    \begin{axis}[
        width=7cm,
        height=5.5cm,
        xlabel={$\maxN^{-1}$},
        ylabel={\texttt{freeAmp}},ylabel shift = -0.5em,
        legend cell align=left,
        legend style={
            at={(1.02,0.5)},
            anchor=west,nodes={inner sep=0.15cm,text depth=0.0em},
            },
        ymajorgrids=true,
        grid style=dashed,
        scatter/classes={
            0={mark=*, plot1},
            4={mark=diamond*, plot2},
            8={mark=square*, plot3},
            12={mark=triangle*, plot4}
        },
        xmin=0,
        xmax=0.055,
        ymin=0,
        scaled x ticks=false,
        xticklabel style={
        /pgf/number format/fixed},
    ]
    \addlegendimage{empty legend}
    \addplot[
        only marks,
        scatter,
        mark size=1.5pt,scatter src=explicit]
    table[
        x="oneOverN",
        y="freeImp",
        meta="nMinusJ",
        col sep=comma]
    {data/singlePoleData.csv};
    \legend{{\hspace{-0.5cm}\maxJ-\maxN:}, 0, 4, 8, 12}
\end{axis}
\end{tikzpicture}\hspace{-2cm}
    \centering
    \caption{In contrast to amplitude maximizing 3-point coupling found in \cite{smatrix3}, the one minimizing a-anomaly doesn't saturate bounds on threshold term. The improvement term in $\tilde {\mathcal T}_{BB\to BB}$ doesn't show any convergence pattern.}
    \label{singleimprov}

\end{figure}
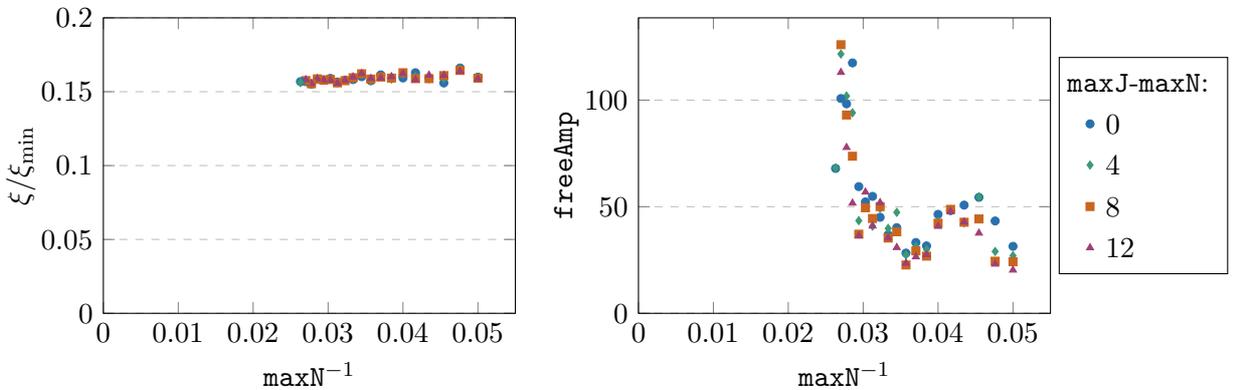

On the other hand, one would expect that the parameter associated with bound state at threshold, $\frac{1}{\rhoVar{s}{4/3}}$ would be saturating analytic bounds, as in scalar theories investigated in \cite{smatrix3}. This is not the case, as optimized amplitude has associated parameter \texttt{boundState} converging to about 0.15 of minimal value of $-32\sqrt 6 \pi$, as shown on a plot. The second of improvement terms, \texttt{fAmp} doesn't appear to converge to any value, similarly to elements of $\gamma_{abc}$, as expected with having an ansatz that is (to some extent) redundant. Both patterns are shown on figure \ref{singleimprov}.

\clearpage

\subsection{A story of two particles}

When considering resonances coming from exchange of massive particle $X$ of mass $m_X^2 > 1$, the picture changes drastically. With square of mass  $m_X$ close to 1 or 4 the absolute minimum of $a$-anomaly is close to $0.15\cdot\aFree$ found which was expected, as these contributions are similar either to pole at $s=1$ or to the threshold singularity. However, in between, the minimal value of anomaly dips at $a \approx 0.034\cdot \aFree$ for the mass $m_X^2 = 2.5\pm0.1$.

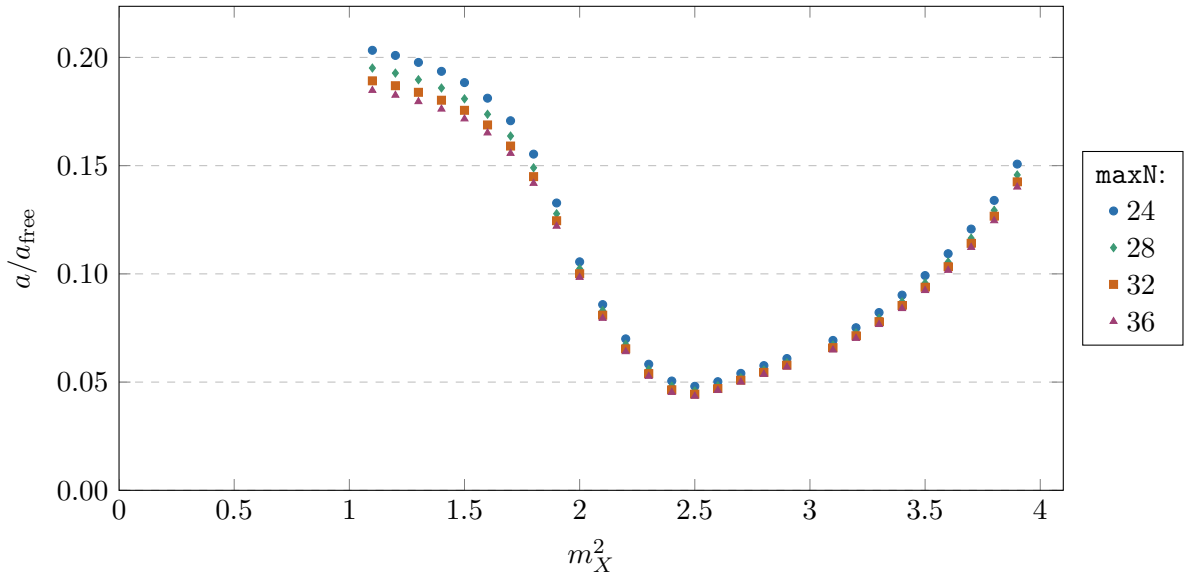
\begin{figure}[H]
    \begin{tikzpicture}
    \begin{axis}[
        width=14cm,
        height=8cm,
        xlabel={$m_X^2$},
        ylabel={$a/ \aFree$},
        legend cell align=left,
        legend style={
            at={(1.02,0.5)},
            anchor=west,nodes={inner sep=0.12cm,text depth=0.0cm},
            },
        ymajorgrids=true,
        grid style=dashed,
        scatter/classes={
            24={mark=*, plot1},
            28={mark=diamond*, plot2},
            32={mark=square*, plot3},
            36={mark=triangle*, plot4}
        },
        ymin=0,
        xmin=0,
        xmax=4.1,
        scaled ticks=false,
        yticklabel style={
        /pgf/number format/fixed,
        /pgf/number format/fixed zerofill,
        /pgf/number format/precision=2},
        xticklabel style={
        /pgf/number format/fixed}
    ]

    \addlegendimage{empty legend}
    \addplot[
        only marks,
        scatter,
        mark size=1.5pt,scatter src=explicit]
    table[
        x="mX2",
        y="aAnomaly",
        meta="maxN",
        col sep=comma]
    {data/doublePoleData.csv};
    \legend{{\hspace{-0.4cm}\maxN:}, 24, 28, 32, 36}
\end{axis}
\end{tikzpicture}\vspace{-0.8cm}
    \centering
    \caption{Absolute minimum of $a$-anomaly as a function of the mass square of the second particle. For each data point $\maxJ = \maxN + 12$}
    \label{doublepoleconvergence}
\end{figure}

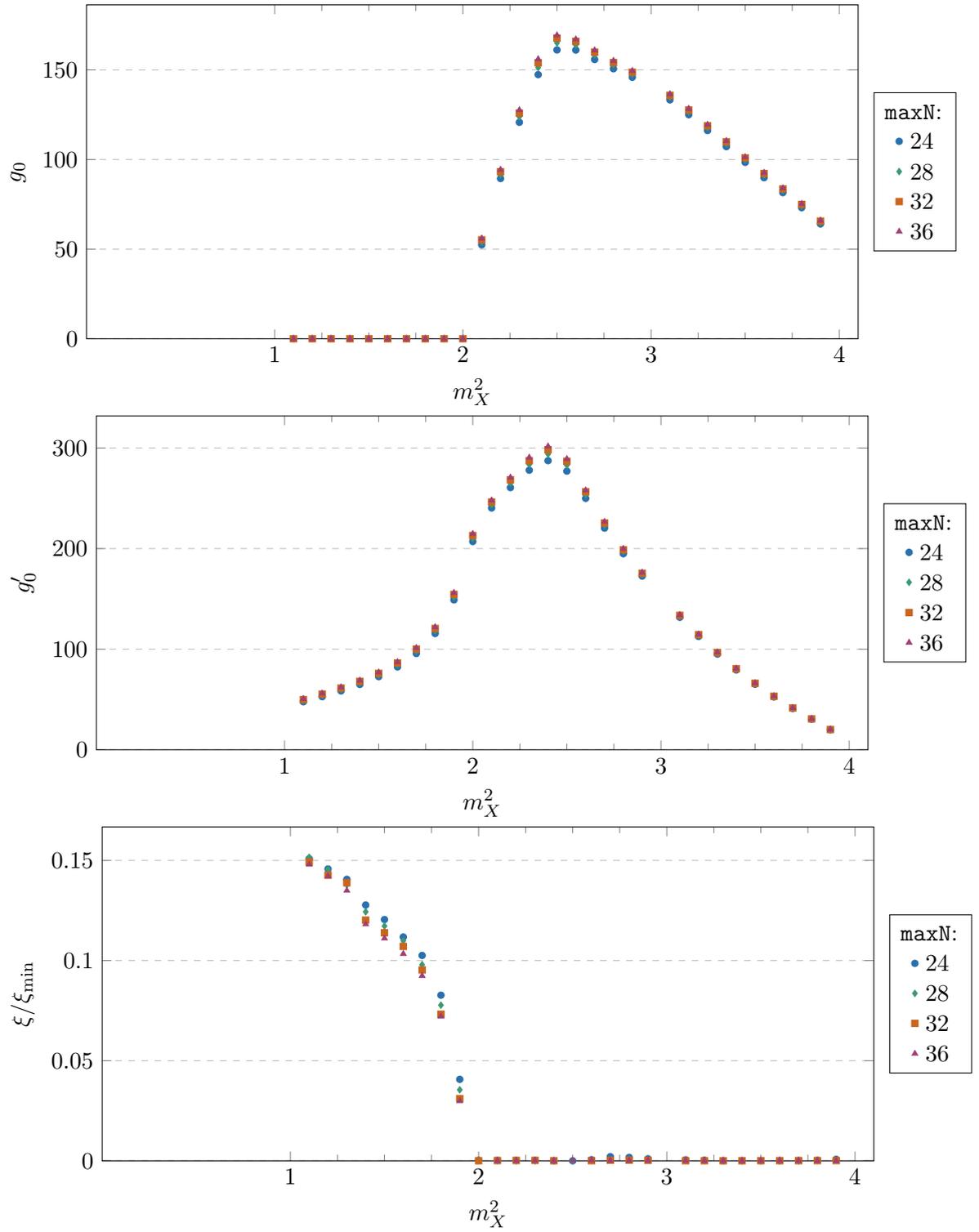
\begin{figure}[ht]
    \begin{tikzpicture}
    \begin{axis}[
        width=14cm,
        height=7cm,
        xlabel={$m_X^2$},
        ylabel={$g_0$},
        legend cell align=left,
        legend style={
            at={(1.02,0.5)},
            anchor=west,nodes={inner sep=0.12cm,text depth=0.0cm},
            },
        ymajorgrids=true,
        grid style=dashed,
        scatter/classes={
            24={mark=*, plot1},
            28={mark=diamond*, plot2},
            32={mark=square*, plot3},
            36={mark=triangle*, plot4}
        },
        ymin=0,
        xmin=0,
        xmax=4.1,
        scaled ticks=false,
        yticklabel style={
        /pgf/number format/fixed},
        xticklabel style={
        /pgf/number format/fixed},
        xtick={1,2,3,4},
        minor x tick num=3
    ]

    \addlegendimage{empty legend}
    \addplot[
        only marks,
        scatter,
        mark size=1.5pt,scatter src=explicit]
    table[
        x="mX2",
        y="g0",
        meta="maxN",
        col sep=comma]
    {data/doublePoleData.csv};
    \legend{{\hspace{-0.4cm}\maxN:}, 24, 28, 32, 36}
\end{axis}
\end{tikzpicture}
    \begin{tikzpicture}
    \begin{axis}[
        width=14cm,
        height=7cm,
        xlabel={$m_X^2$},
        ylabel={$g'_0$},
        legend cell align=left,
        legend style={
            at={(1.02,0.5)},
            anchor=west,nodes={inner sep=0.12cm,text depth=0.0cm},
            },
        ymajorgrids=true,
        grid style=dashed,
        scatter/classes={
            24={mark=*, plot1},
            28={mark=diamond*, plot2},
            32={mark=square*, plot3},
            36={mark=triangle*, plot4}
        },
        ymin=0,
        xmin=0,
        xmax=4.1,
        scaled ticks=false,
        yticklabel style={
        /pgf/number format/fixed},
        xticklabel style={
        /pgf/number format/fixed},
        xtick={1,2,3,4},
        minor x tick num=3
    ]

    \addlegendimage{empty legend}
    \addplot[
        only marks,
        scatter,
        mark size=1.5pt,scatter src=explicit]
    table[
        x="mX2",
        y="g0X",
        meta="maxN",
        col sep=comma]
    {data/doublePoleData.csv};
    \legend{{\hspace{-0.4cm}\maxN:}, 24, 28, 32, 36}
\end{axis}
\end{tikzpicture}
    \begin{tikzpicture}
    \begin{axis}[
        width=14cm,
        height=7cm,
        xlabel={$m_X^2$},
        ylabel={$\xi/ \xi_\textrm{min}$},
        legend cell align=left,
        legend style={
            at={(1.02,0.5)},
            anchor=west,nodes={inner sep=0.12cm,text depth=0.0cm},
            },
        ymajorgrids=true,
        grid style=dashed,
        scatter/classes={
            24={mark=*, plot1},
            28={mark=diamond*, plot2},
            32={mark=square*, plot3},
            36={mark=triangle*, plot4}
        },
        ymin=0,
        xmin=0,
        xmax=4.1,
        scaled ticks=false,
        yticklabel style={
        /pgf/number format/fixed},
        xticklabel style={
        /pgf/number format/fixed},
        xtick={1,2,3,4},
        minor x tick num=3
    ]

    \addlegendimage{empty legend}
    \addplot[
        only marks,
        scatter,
        mark size=1.5pt,scatter src=explicit]
    table[
        x="mX2",
        y="xi",
        meta="maxN",
        col sep=comma]
    {data/doublePoleData.csv};
    \legend{{\hspace{-0.4cm}\maxN:}, 24, 28, 32, 36}
\end{axis}
\end{tikzpicture}
    \centering
    \caption{Size of 3-point couplings $g_0$ and $g'_0$ and threshold singularity $\xi$ with respect to additional particle mass $m_X^2$}
    \label{doublepolepattern}
\end{figure}

The behavior of three-point couplings (shown on figure \ref{doublepolepattern}) and threshold singularity term can be separated into two regions. For $m_X^2 \le 2$ the value of $g_0$ is close to 0, and $|g_0'|$ (matter-matter-$X$) saturates the unitarity bound of 3-point coupling found in previous S-matrix bootstrap experiments\cite{smatrix3}. Above $m_X^2 = 2$ the coupling $|g_0|$ starts to grow rapidly, maximizing around $m_X^2 = 2.5\pm0.1$. The opposite applies to threshold singularity term $\xi$ - it decays quickly from $\approx 0.15$ of minimal value coming from unitarity bounds ($\xi_\textrm{min}= -32 \sqrt{6}\pi$) to decay to 0 at $m_X^2 = 2$. It looks like a pole below $m_X^2 = 2$ absolutely consumes pole at $m_A^2=1$, and pole above consumes threshold singularity, when it comes to minimization of $a$-anomaly.

\begin{figure}[ht]
    \hspace{-2.5cm}%
    \begin{tikzpicture}
    \begin{axis}[
        width=7cm,
        height=5.5cm,
        xlabel={$m_X^2$},
        ylabel={$g_2$},
        legend cell align=left,
        legend style={
            at={(1.02,0.5)},
            anchor=west,nodes={inner sep=0.12cm,text depth=0.0cm},
            },
        ymajorgrids=true,
        grid style=dashed,
        scatter/classes={
            24={mark=*, plot1},
            28={mark=diamond*, plot2},
            32={mark=square*, plot3},
            36={mark=triangle*, plot4}
        },
        ymin=0,
        xmin=0,
        xmax=4.1,
        scaled x ticks=false,
        yticklabel style={
        /pgf/number format/fixed},
        xticklabel style={
        /pgf/number format/fixed},
        xtick={1,2,3,4},
        minor x tick num=3
    ]

    \addlegendimage{empty legend}
    \addplot[
        only marks,
        scatter,
        mark size=1.5pt,scatter src=explicit]
    table[
        x="mX2",
        y="g2",
        meta="maxN",
        col sep=comma]
    {data/doublePoleData.csv};
\end{axis}
\end{tikzpicture}%
    \begin{tikzpicture}
    \begin{axis}[
        width=7cm,
        height=5.5cm,
        xlabel={$m_X^2$},
        ylabel={$g_2'$},
        legend cell align=left,
        legend style={
            at={(1.02,0.5)},
            anchor=west,nodes={inner sep=0.12cm,text depth=0.0cm},
            },
        ymajorgrids=true,
        grid style=dashed,
        scatter/classes={
            24={mark=*, plot1},
            28={mark=diamond*, plot2},
            32={mark=square*, plot3},
            36={mark=triangle*, plot4}
        },
        ymin=0,
        xmin=0,
        xmax=4.1,
        scaled x ticks=false,
        yticklabel style={
        /pgf/number format/fixed,
        /pgf/number format/precision=10},
        xticklabel style={
        /pgf/number format/fixed},
        xtick={1,2,3,4},
        minor x tick num=3
    ]

    \addlegendimage{empty legend}
    \addplot[
        only marks,
        scatter,
        mark size=1.5pt,scatter src=explicit]
    table[
        x="mX2",
        y="g2X",
        meta="maxN",
        col sep=comma]
    {data/doublePoleData.csv};
    \legend{{\hspace{-0.4cm}\maxN:}, 24, 28, 32, 36}
\end{axis}
\end{tikzpicture}%
    \hspace{-2cm}
    \centering\vspace{-0.2cm}
    \caption{Size of 3-point couplings $g_2$ and $g'_2$ with respect to additional particle mass $m_X^2$. Note the kink of $g'_2$ around $m_X^2 = 2$ -- it is very likely similar discontinuity applies to $g'_0$, however it may be difficult to notice it due to low resolution in $m_X^2$.}
    \label{doublepolepattern2}
\end{figure}
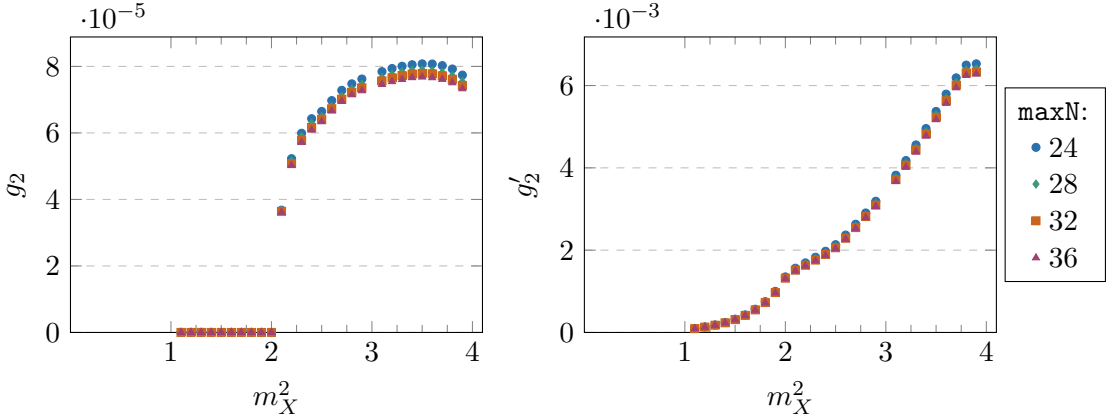

Somewhat similar behavior (of disappearance of resonance at mass of $A$) is observed at dilaton-dilaton-matter and dilaton-dilaton-X couplings, as plotted in figure \ref{doublepolepattern2}.

\clearpage

\subsection{A story of many particles}

\begin{figure}[H]
    \begin{tikzpicture}
    \begin{axis}[
        width=14cm,
        height=6.5cm,
        xlabel={$\maxN^{-1}$},
        ylabel={$a/ \aFree$},
        legend cell align=left,
        legend style={
            at={(1.02,0.5)},
            anchor=west,nodes={inner sep=0.1cm,text depth=0.0em},
            },
        ymajorgrids=true,
        grid style=dashed,
        xmin=0.0,
        xmax=0.055,
        ymin=0.02,
        ymax=0.06,
        scaled ticks=false,
        xticklabel style={
            /pgf/number format/fixed,
            /pgf/number format/precision=10},
        yticklabel style={
        /pgf/number format/fixed}
    ]

    \addlegendimage{empty legend}
    \addplot[
        only marks,
        scatter,
        mark size=1.5pt]
    table[
        x="oneOverN",
        y="aAnomaly",
        point meta=1,
        col sep=comma]
    {data/multipole_convergence.csv};
    \addplot[domain=0:0.055, dashed]{0.0358698+0.2733459*x};
\end{axis}
\end{tikzpicture}%
    \centering\vspace{-0.2cm}
    \caption{Convergence of $a$-anomaly bounds with multiple poles allowed. $\maxJ = \maxN + 12$ for all data points.}
    \label{multipoleconvergence}
\end{figure}
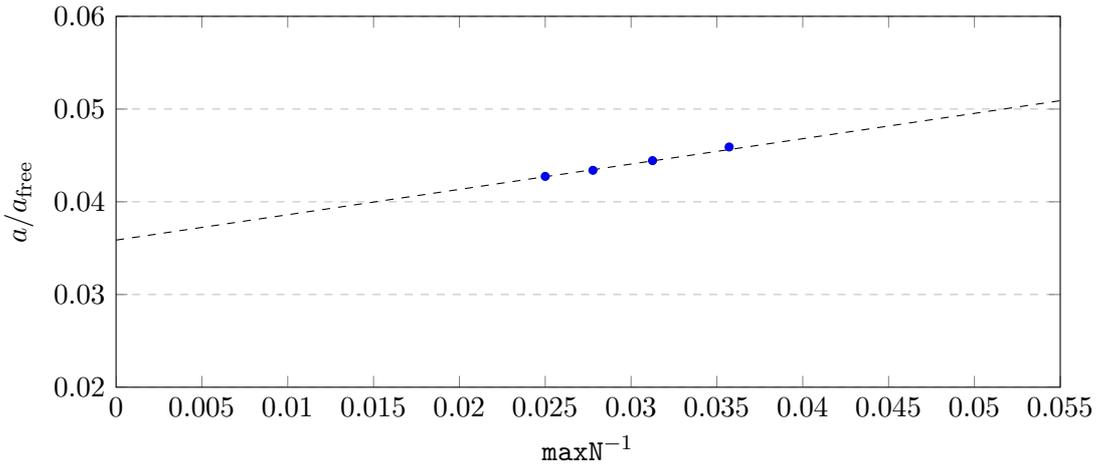

To approach a general theory containing \textit{at least} a single particle of spin 0, a theory with many possible resonances is investigated. Instead of introducing single extra resonance of mass $m_X$, a set of them is included in the ansatz, with masses\footnote{The resonance at $m_i = 3.0$ is missing due to a numerical error.}
\begin{equation}
    m_i^2 = 1.1, 1.2, \dots, 3.8, 3.9
\end{equation}
with respective 3-point couplings $g_0\!\left(m_i\right)$ and $g_2\!\left(m_i\right)$, with $A$'s and with dilatons respectively.
Surprisingly, the absolute value of $a$-anomaly in such case can be extrapolated to
\begin{equation}
    a_\textrm{min} \approx 0.036 \cdot \aFree
\end{equation}
with convergence pattern presented on the figure \ref{multipoleconvergence}.

The exact numerical result shall be taken with a grain of salt, as little data points were evaluated due to high complexity of computations.

However, this value being so close to minimum of $a$-anomaly from previous section shall bring attention to 3-point couplings related to each of allowed intermediate particles.
\begin{figure}[ht]
    \begin{tikzpicture}
    \begin{semilogyaxis}[
        width=14cm,
        height=8cm,
        xlabel={$m_X^2$},
        ylabel={$g_0\!\left(m_i\right)$},
        legend cell align=left,
        legend style={
            at={(1.02,0.5)},
            anchor=west,nodes={inner sep=0.12cm,text depth=0.0cm},
            },
        ymajorgrids=true,
        grid style=dashed,
        scatter/classes={
            28={mark=*, plot1},
            32={mark=diamond*, plot2},
            36={mark=square*, plot3},
            40={mark=triangle*, plot4}
        },
        xmin=0,
        xmax=4.1,
        scaled ticks=false,
        yticklabel style={
        /pgf/number format/fixed},
        xticklabel style={
        /pgf/number format/fixed},
        xtick={1,2,3,4},
        minor x tick num=3
    ]
    \addlegendimage{empty legend}
    \addplot[
        only marks,
        scatter,
        mark size=1.5pt,scatter src=explicit]
    table[
        x="m",
        y="g0",
        meta="maxN",
        col sep=comma]
    {data/multipole_matter.csv};
    \legend{{\hspace{-0.4cm}\maxN:}, 28,32,36,40}
\end{semilogyaxis}
\end{tikzpicture}
    \begin{tikzpicture}
    \begin{semilogyaxis}[
        width=14cm,
        height=8cm,
        xlabel={$m_X^2$},
        ylabel={$g_2\!\left(m_i\right)$},
        legend cell align=left,
        legend style={
            at={(1.02,0.5)},
            anchor=west,nodes={inner sep=0.12cm,text depth=0.0cm},
            },
        ymajorgrids=true,
        grid style=dashed,
        scatter/classes={
            28={mark=*, plot1},
            32={mark=diamond*, plot2},
            36={mark=square*, plot3},
            40={mark=triangle*, plot4}
        },
        xmin=0,
        xmax=4.1,
        scaled ticks=false,
        yticklabel style={
        /pgf/number format/fixed},
        xticklabel style={
        /pgf/number format/fixed},
        xtick={1,2,3,4},
        minor x tick num=3
    ]

    \addlegendimage{empty legend}
    \addplot[
        only marks,
        scatter,
        mark size=1.5pt,scatter src=explicit]
    table[
        x="m",
        y="g2",
        meta="maxN",
        col sep=comma]
    {data/multipole_dilaton.csv};
    \legend{{\hspace{-0.4cm}\maxN:}, 28,32,36,40}
\end{semilogyaxis}
\end{tikzpicture}\vspace{-0.6cm}
    \centering
    \caption{Relative size of couplings corresponding to each pole. Note the logarithmic scale.}
    \label{multipolecouplings}
\end{figure}
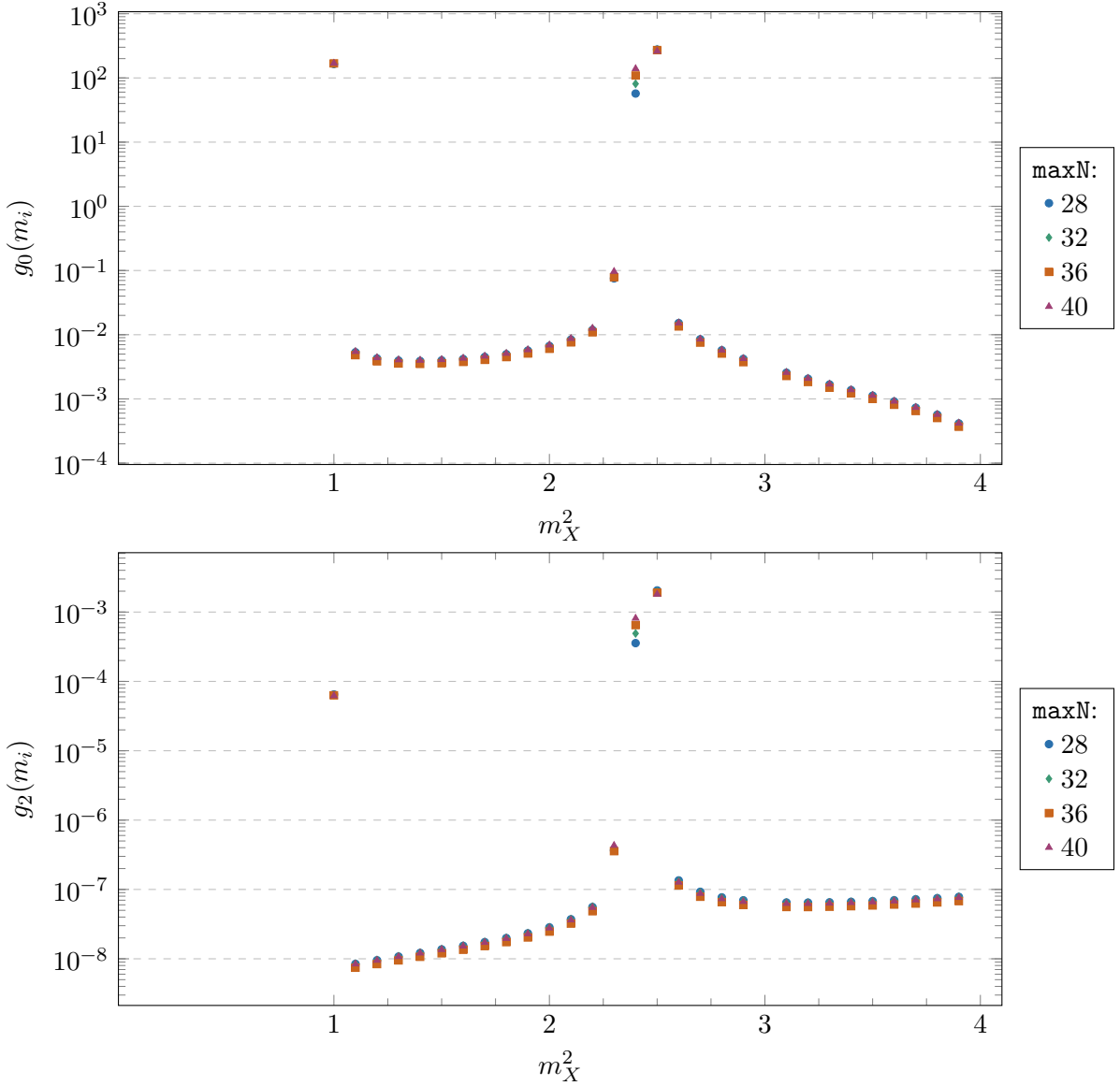

When taking a closer look on figure \ref{multipolecouplings}, one can notice the only significant contribution to the amplitudes results from resonances at mass $m_i^2 = 2.4, m_i=2.5$. The natural conjecture from this observation is that the non-trivial theory that really minimizes $a$-anomaly contains two stable particles, one of mass $m_A$ (normalized to 1 in the experiment), and another with $m_X^2$ between 2.4 and 2.5. The further investigation, with more detailed grid of allowed resonances, shall eventually bring a definitive answer to question of minimal $a$-anomaly of non-trivial theory containing at least one spin-0 particle.
\section{Discussion}\label{discussion}

The question of absolute minimum of $a$-anomaly of UV CFTs is far from answered. The research presented in this paper finds (finite, non-zero) bound on $a$-anomaly for range of theories including stable scalars, and with results, more questions arise.

The data on figure \ref{multipolecouplings} suggest the scalar theory minimizing $a$ maybe consists only of two particles, with mass related by $m_X^2 \approx 2.4 m_A^2$, and more precise numerical experiments in terms of allowed masses $m_X$ shall be considered. The possibility of existence of theories with higher-spin matter can neither be ruled out.

However, none of such experiments can answer the question "What is this ($a$-minimizing) theory?", or what is the UV CFT corresponding to it. For now, the answer is unknown.
\section{Acknowledgements}
{\small The author is supported by the Swiss National Science Foundation through the project
200020$\_$197160 and through the National Centre of Competence in Research SwissMAP.

The author also received some support from the Simons Foundation grant 488649 (Simons Collaboration on the Non-
perturbative Bootstrap).

The author wants to thank Biswajit Sahoo, Denis Karateev and João Penedones for countless fruitful discussions, time, and guidance regarding the research presented in this paper.}

\bibliographystyle{JHEP}
\bibliography{main}

\end{document}